\documentclass[sigconf]{acmart}
\usepackage{lipsum}
\usepackage{listings}
\usepackage{xcolor}
\usepackage{hyperref}
\setcitestyle{numbers}
\usepackage{tcolorbox}
\usepackage{enumitem}
\usepackage{booktabs}
\usepackage{float}
\usepackage{soul}
\usepackage{xcolor}
\sethlcolor{cyan!30}

\tcbset{
  colback=gray!10, 
  colframe=gray!50, 
  boxrule=0.5pt,
  arc=2pt,
  left=6pt,
  right=6pt,
  top=4pt,
  bottom=4pt
}

\acmConference[2026]{2026: Under Review}{April 2026}{xyz, arxiV}

\usepackage{graphicx}

\title{Cluster-guided LLM-Based Anonymization of Software Analytics Data: Studying Privacy-Utility Trade-offs in JIT Defect Prediction
}

\author{Maaz Khan}
\email{maaz.sfg@gmail.com}
\affiliation{
  \institution{SFG Lab, Lahore University of Management Sciences}
  \city{Lahore}
  \country{Pakistan}
}

\author{Gul Sher Khan Swati}
\email{gulsher.sfg@gmail.com}
\affiliation{
  \institution{SFG Lab, Lahore University of Management Sciences}
  \city{Lahore}
  \country{Pakistan}
}

\author{Ahsan Raza}
\email{ahsan.sfg@gmail.ca}
\affiliation{
  \institution{SFG Lab, Lahore University of Management Sciences}
  \city{Lahore}
  \country{Pakistan}
}

\author{Pir Sami Ullah Shah}
\email{samiullah.shah@nu.edu.pk}
\affiliation{
  \institution{National University of Computer and Emerging Sciences (FAST)}
  \city{Islamabad}
  \country{Pakistan}
}

\author{Abdul Ali Bangash}
\email{abdulali@lums.edu.pk}
\affiliation{
  \institution{SFG Lab, Lahore University of Management Sciences}
  \city{Lahore}
  \country{Pakistan}
}

\date{October 2025}

\begin{document}

\begin{abstract}

The increasing use of machine learning (ML) for Just-In-Time (JIT) defect prediction raises concerns about privacy leakage from software analytics data. Existing anonymization methods—such as tabular transformations and graph perturbations—often overlook contextual dependencies among software metrics, leading to suboptimal privacy–utility trade-offs. Leveraging the contextual reasoning of Large Language Models (LLMs), we propose a cluster-guided anonymization technique that preserves contextual and statistical relationships within JIT datasets. Our method groups commits into feature-based clusters and employs an LLM to generate context-aware parameter configurations for each commit cluster, defining $\alpha-\beta$ ratios and churn mixture distributions used for anonymization. Our evaluation on six projects (\textit{Cassandra}, \textit{Flink}, \textit{Groovy}, \textit{Ignite}, \textit{OpenStack}, and \textit{Qt}) shows that our LLM-based approach achieves privacy level~2 (IPR~$\geq$~80\%), improving privacy by 18–25\% over four State of the Art Graph-based anonymization baslines while maintaining comparable F1-scores. Our results demonstrate that LLMs can act as adaptive anonymization engines when we provide them with cluster-specific statistical information about similar data points, enabling context-sensitive and privacy-preserving software analytics without compromising predictive accuracy.

\end{abstract}

\settopmatter{printacmref=false}

\maketitle

\section{Introduction}
Just-In-Time (JIT) defect prediction is a critical technique in software engineering, enabling developers to identify and mitigate defects at the commit level, thereby reducing the time and effort required for quality assurance~\cite{kamei2013large,hoang2019deepjit}. Unlike traditional module-level defect prediction, JIT models provide timely feedback on code changes, allowing teams to prioritize resources on risky commits and improve overall software reliability~\cite{pornprasit2021jitline,bryan2023graph}. This approach has been widely studied in the Mining Software Repositories (MSR) community, with empirical evaluations demonstrating its effectiveness across large-scale datasets from open-source projects~\cite{keshavarz2022apachejit,wang2023just}. However, advancing JIT defect prediction research relies heavily on the availability of shared software analytics data, such as commit histories, metrics, and defect labels, to facilitate replication, benchmarking, and cross-project learning~\cite{peters2013balancing}. Despite its benefits, the sharing of software analytics data poses significant privacy risks, as it often reveals sensitive information about development processes, intellectual property, or individual contributors~\cite{morales2019systematic,herzig2016impact}. It has been demonstrated that the training data of machine learning (ML) models, including those used for defect prediction, can be reconstructed through adversarial techniques such as shadow modeling~\cite{shokri2017membership,wang2022reconstructing}. These vulnerabilities underscore the need for robust privacy-preserving mechanisms in software data sharing.

Prior work has explored data anonymization to address the privacy-utility trade-off in defect prediction datasets. Tabular\\ anonymization techniques, such as generalization, perturbation, and pseudonymization, have been applied to structured data like commit metrics, aiming to mask identifiers while retaining statistical properties~\cite{fung2010privacy}. In the context of software defect prediction, specific methods like MORPH, which mutates data instances by perturbing attribute values to obfuscate sensitive information while attempting to preserve decision boundaries, and LACE/LACE2, which integrate data minimization (e.g., via instance selection) with localized perturbation to enforce closeness among similar records, have been developed to enable secure cross-project data sharing~\cite{peters2013balancing,peters2012privacy,peters2015lace2}. Empirical evidence from cross-company defect prediction studies shows that MORPH provides substantial privacy gains but can reduce predictive performance in some cases, whereas LACE2 improves upon MORPH by achieving higher utility while offering comparable or better privacy levels, as demonstrated in evaluations across multiple open-source datasets~\cite{peters2012privacy}. More advanced approaches treat software repositories as knowledge graphs, employing graph-based anonymization methods (e.g., k-degree anonymity) to preserve relational structures. For example, studies have investigated random add/delete, random switch, k-DA, and generalization techniques on JIT datasets to evaluate their impact on privacy scores and model performance~\cite{malik2024towards}. While these methods effectively balance privacy and utility in terms of structural and statistical fidelity, they often overlook the contextual nuances of the data.

In this paper, we introduce an approach that leverages Large Language Models to anonymize Just-In-Time defect prediction datasets~\cite{keshavarz2022apachejit,mcintosh2018fix}. Unlike traditional methods that apply generic anonymization techniques, our approach utilizes LLMs to preserve the contextual integrity of cluster-specific data distributions. Because LLMs can model semantic and structural dependencies~\cite{cirillo2025augmenting}, they may help enable anonymization methods that better preserve both privacy and utility. This context-aware capability establishes a new rationale for using LLMs in anonymization: they can understand and preserve data semantics while performing privacy transformation, offering a more effective solution compared to existing graph-based methods. To support transparency and reproducibility, we have released all our datasets, experimental scripts, and analysis code under an open license at the following link.\footnote{Replication package available at \url{https://anonymous.4open.science/r/anonymizer-27A6}, containing all code, scripts, datasets, and results required to fully reproduce this study.}

We initially perform a preliminary study to exploit whether Large Language Models—specifically Grok 4 and DeepSeek V3—could anonymize JIT defect prediction datasets through zero-shot prompting. Using commit data from Apache Cassandra project~\cite{keshavarz2022apachejit}, we incrementally scaled the input from 50 to 2,000 commits and observed that the LLMs could produce syntactically coherent and contextually consistent anonymizations. However, while our experiments indicated that LLMs hold promise for privacy-preserving software analytics, they also revealed limitations in preserving statistical distributions and ensuring consistent anonymization behavior as dataset size increased. 

Building on these insights, our \textbf{cluster-guided LLM-based anonymization technique} introduces a structured, multi-step process that integrates statistical grouping with context-aware parameter generation. First, we organize commits into feature-based clusters using quantile binning of quasi-identifiers (e.g., \texttt{ndev}, \texttt{nf}, \texttt{ent}), capturing groups of commits with similar change behaviors. For each cluster, we compute descriptive statistics of code change metrics such as lines added (\texttt{la}), lines deleted (\texttt{ld}), and churn. We then provide these statistical profiles to the LLM, which generates anonymization parameters, including mixture distributions for churn and ratio constraints, that guide the regeneration of anonymized values. By prompting the model with cluster-level statistical summaries instead of raw records, our approach enables the LLM to infer realistic yet privacy-preserving transformations aligned with each cluster’s statistical context. Our evaluation across six JIT defect prediction datasets shows that this method achieves substantially stronger privacy (up to 90.2\% IPR, a +18--25\% improvement over graph-based baselines) while maintaining comparable predictive utility. This establishes a new paradigm where LLMs, when provided with the statistical context of the cluster, can act as adaptive anonymization engines capable of preserving both statistical structure and contextual integrity in software analytics data.

\paragraph{\textbf{Case Study: Privacy-Preserving Data Sharing in a Software Company.}}
Consider a mid-sized software company that wants to collaborate with universities by releasing its commit-level dataset for defect-prediction research. The dataset contains sensitive metrics such as lines added (la) and lines deleted (ld). Publishing these raw values can reveal operational patterns ~\cite{peters2012privacy}, for example, a developer consistently adding 100–150 lines per day, or a team performing unusually large churn spikes before release deadlines. Even if an external party already knows the overall distribution of code changes (e.g., that most commits fall between 50–200 churn lines), the specific per-commit patterns can still expose confidential information such as team productivity rhythms, code-review bottlenecks, or bug-fixing workload cycles. Such instance-level patterns may be exploited by competitors to infer staffing levels or by clients during negotiations to question development speed.


\begin{enumerate}
    \item \textbf{Proposing a novel clustering-based representation} of software analytics data, in this case JIT defect prediction datasets, where commits are grouped into statistically similar clusters through quantile-based binning of quasi-identifiers. 
    
    \item \textbf{Proposing a methodology to generate sensitive attributes for anonymization}, where cluster-level statistical summaries (e.g., mean, quantiles and range) are used to guide LLMs in producing parameter configurations (e.g churn mixture and beta ratio) that are used to generate anonymized values.
    
    \item \textbf{Evaluating the proposed LLM-based anonymization approach} against four graph-based anonymization baselines to assess its effectiveness in achieving stronger privacy–utility balance in JIT defect prediction data.
\end{enumerate}

To evaluate the effectiveness of our approach, we address the following research questions.

\begin{description}[leftmargin=3.2em, style=sameline]
  \item \textbf{RQ1:} Can LLMs directly anonymize JIT defect metrics through zero-shot prompting while maintaining privacy and model utility?
  \item \textbf{RQ2:} How does our cluster-guided LLM-based technique compare with graph anonymization on privacy and utility?
  \item \textbf{RQ3:} Do these privacy and utility outcomes hold across different LLMs?
\end{description}

\section{Background and Related Work}
This section reviews prior work on anonymizing software defect prediction data, particularly for Just-In-Time (JIT) prediction. 

\subsection{Just-In-Time Defect Prediction}
Just-In-Time (JIT) defect prediction aims to identify defect-inducing commits at the moment of code submission, providing developers with immediate feedback that improves software quality~\cite{shahini2025calibration,yu2024formal}.
Unlike traditional defect-prediction approaches that operate at the module or file level, JIT models analyze changes at the commit level, enabling finer-grained insights into risky modifications~\cite{shahini2024empirical,xu2024jitsmart}.
JIT datasets typically contain commit-level metrics that describe both the magnitude of a change and the developer activity involved, including lines added (la), lines deleted (ld), code churn, number of files changed (nf), number of developers (ndev), and entropy (ent)~\cite{zhao2023systematic}.

Large-scale open-source projects such as Apache Cassandra, Flink, Groovy, Ignite, OpenStack, and Qt are frequently used as benchmarks in the Mining Software Repositories (MSR) community~\cite{keshavarz2022apachejit,shahini2024empirical,mcintosh2018fix}.
These datasets, consolidated in corpora such as ApacheJIT~\cite{keshavarz2022apachejit}, enable replication, benchmarking, and cross-project learning, forming the empirical foundation for much of the progress in software-quality modeling~\cite{shahini2025calibration,xu2024jitsmart}.

\subsection{Privacy Risks in Software Analytics Data}
Despite their utility, commit-level datasets often contain sensitive organizational information. The metrics used in defect prediction can reveal operational and performance details such as development velocity, code-review efficiency, or bug-resolution frequency~\cite{zimmermann2009cross,yamamoto2023towards}. In cross-company contexts, such information can expose how many lines of code are added or deleted daily and how many bugs the company resolves weekly, details that could affect corporate reputation if misinterpreted. Prior research has shown that sharing internal project data raises confidentiality concerns and competitive risks~\cite{weyuker2010assessing,chen2019dpshare}. Moreover, recent studies have demonstrated that the training data of machine-learning models can be reconstructed through adversarial techniques such as shadow modeling and model-inversion attacks~\cite{zhou2023boosting,shokri2024model}. These attacks enable an adversary to approximate or regenerate original datasets from trained models, posing a significant privacy threat even when raw data is not explicitly shared~\cite{du2020towards}. Such vulnerabilities underscore the urgent need for robust, privacy-preserving mechanisms in software-engineering data sharing.

\subsection{Traditional Anonymization Approaches}

To address privacy risks in defect prediction datasets, various \\anonymization techniques have been proposed to modify sensitive attributes while retaining their utility for training predictive models. 
Tabular approaches such as MORPH, LACE, and LACE2 perturb numeric metrics or selectively minimize data instances to balance privacy and model performance~\cite{peters2012privacy,peters2015lace2}. 
MORPH introduces controlled noise into sensitive attributes such as lines added and deleted, whereas LACE2 enhances this by combining data minimization with neighborhood-based perturbation to improve statistical similarity between original and anonymized data. 
Although these approaches achieve a trade-off between privacy and utility, stronger anonymization often reduces predictive accuracy. 
Moreover, they treat each attribute independently, ignoring the semantic relationships among metrics. 
For instance, in JIT defect prediction, the number of lines added is contextually related to the number of files modified and overall churn, yet traditional tabular methods anonymize these metrics in isolation. 
This neglect of inter-feature dependencies leads to unrealistic or incoherent anonymized data that can misguide downstream models.

To address this limitation, recent work has explored graph-based anonymization techniques such as \textit{k}-Degree Anonymity (k-DA), Random Add/Delete (RAD), and Random Switch (RS)~\cite{malik2024towards}. 
These methods represent software repositories as graphs where nodes correspond to entities (e.g., commits, developers, files) and edges capture their relationships. 
Anonymization is achieved by modifying the graph’s structure—adding, deleting, or switching edges—to obscure identifiable patterns while preserving overall degree distributions or topological properties. 
Although graph-based approaches better capture relational structures within software projects, they still lack contextual awareness. 
Thus, while they anonymize structural relationships, they fail to preserve the contextual dependencies and interpretability essential for reliable defect prediction.

\subsection{Large Language Models for Anonymization}
Recent advances in Large Language Models (LLMs) have opened new possibilities for context-aware anonymization. Unlike traditional methods that depend on fixed statistical masking or noise injection, LLMs can infer semantic relationships and inter-feature dependencies in data~\cite{staab2024large,cirillo2025augmenting}. By understanding both syntactic and statistical patterns, they perform adaptive transformations that protect privacy while preserving data utility.  

Staab et al.~\cite{staab2024large} showed that LLMs outperform conventional anonymization by dynamically balancing privacy and utility. Li et al.~\cite{yan2025protecting} reviewed privacy-preserving mechanisms in LLMs on textual data, emphasizing their ability to maintain contextual coherence. Cirillo et al.~\cite{cirillo2025augmenting} demonstrated that prompt-based anonymization of a tabular financial data, helps retain statistical distributions better than rule-based approaches. Hou et al.~\cite{hou2024large} further noted that LLMs model feature dependencies in structured data, enabling adaptive anonymization.  

By leveraging contextual reasoning, LLM-based techniques generate anonymized data that are coherent, representative, and analytically useful, effectively overcoming the limitations of traditional noise- or constraint-driven approaches. These capabilities position LLMs as a strong foundation for next-generation privacy-preserving data sharing.

\section{Methodology}
This section describes the methodology used to evaluate the effectiveness of LLMs for anonymizing JIT defect prediction data. The experiments are structured around three research questions (RQs): RQ1 – Direct LLM-based anonymization through zero-shot prompting, RQ2 – Cluster-guided LLM based anonymization, and RQ3 – Cross-LLM comparison.

JIT defect prediction seeks to identify software commits that are likely to introduce defects at the moment of change, allowing developers to take preventive action before faults propagate. Each commit is represented by a set of measurable features that describe how much and how broadly the code was modified, such as lines added (\textit{la}), lines deleted (\textit{ld}), code churn (the total amount of change), number of files changed (\textit{nf}), number of developers (\textit{ndev}), and entropy (\textit{ent}), which captures the spread of edits across files. These metrics help predict defect-proneness because large or complex commits (high \textit{la}, \textit{ld}, or \textit{nf}) are statistically more likely to introduce bugs~\cite{kamei2013large}. However, they can also reveal sensitive information about developer activity or organizational productivity. For example, publishing the average number of lines deleted per developer could unintentionally expose team performance patterns~\cite{peters2012privacy}.

To balance usefulness and confidentiality, we categorize metrics into sensitive attributes (e.g., \textit{la}, \textit{ld})~\cite{malik2024towards}, values that can directly disclose internal behaviors, and quasi-identifiers (QIDs) (e.g., \textit{ndev}, \textit{nf}, \textit{ent}), features that are not private on their own but can reveal patterns when combined. The goal of anonymization is therefore to modify the sensitive attributes while preserving the statistical relationship, ensuring that the dataset remains valuable for defect-prediction research without compromising privacy.

\begin{figure*}[t] 
    \centering
    \includegraphics[
        width=\textwidth,
        trim=0 24cm 0 0, 
        clip
    ]{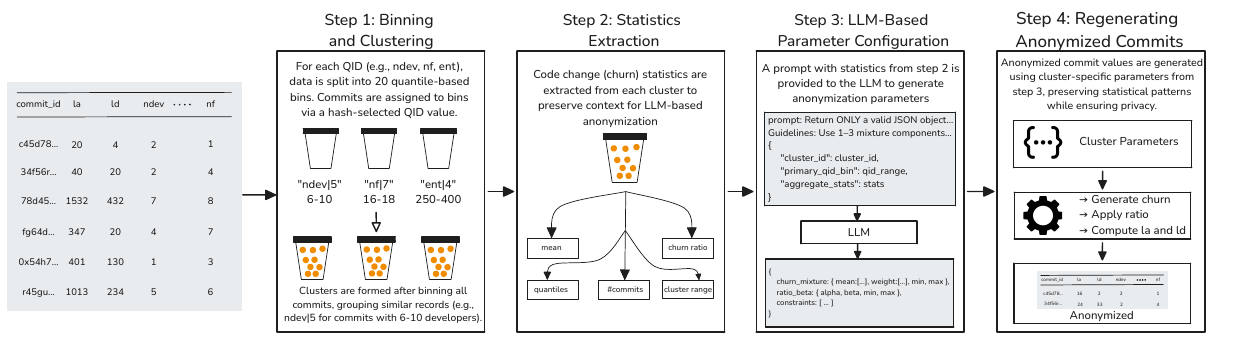}
    
    \caption{The overview of our Cluster-Guided LLM-Based Anonymization approach}
    
    \label{fig:methodology}
\end{figure*}

\subsection{Datasets and Preprocessing}
We used six long-lived open-source projects commonly employed in JIT defect prediction research: Cassandra (6,527 commits), Flink (9,352 commits), Groovy (6,448 commits), Ignite (9,628 commits), OpenStack (10,319 commits), and Qt (20,525 commits)~\cite{keshavarz2022apachejit, mcintosh2018fix, Sliwerski2005WhenDC}. 
Each dataset was released with commit-level defect labels generated by the original authors using the SZZ algorithm~\cite{Sliwerski2005WhenDC}. 
The datasets already include essential process metrics such as lines added (\textit{la}), lines deleted (\textit{ld}), number of developers (\textit{ndev}), and other quasi-identifiers relevant to JIT defect prediction. We computed two additional derived attributes to support privacy and utility evaluation: \textit{churn} = \textit{la} + \textit{ld} and \textit{ratio} = \textit{la} / \textit{churn}.

\subsection{LLM Hyperparameter Settings}
We configured all LLM runs with deterministic decoding settings (\texttt{temperature}=0.0, \texttt{top\_p}=0.0) to produce consistent anonymization outputs and disabled stochastic sampling during text generation to prevent random variation. We repeated every experiment five times under identical conditions and confirmed that the results remained stable, ensuring that outcomes were not due to chance.\cite{Schaefer2023}

\subsection{Preliminary Study}

In order to explore the possibility that LLMs could anonymize commit-level software metrics in JIT defect prediction data without any structural guidance, we examined two models—\textit{Grok 4} and \textit{DeepSeek V3}—on subsets of the ApacheJIT dataset~\cite{keshavarz2022apachejit}, focusing on the Apache Cassandra project. Each model processed the data in sequential batches of 50, 200, 500, 1000, and 2000 commits respectively to analyze how performance scales with input size. Each commit record contained numerical metrics such as \textit{la}, \textit{ld}, \textit{nf}, and \textit{ndev}, which describe the size and complexity of a code change. We prompted the models to modify these values while keeping the overall distribution realistic and internally consistent. For example, a commit with \textit{la} = 120 and \textit{ld} = 60 (moderate change) might be transformed into \textit{la} = 115 and \textit{ld} = 70, preserving proportionality while introducing controlled variation. Smaller commits (e.g., \textit{la} = 8, \textit{ld} = 4) were adjusted with correspondingly minor deviations to retain behavioral plausibility. Both models consistently generated syntactically valid and statistically coherent anonymized data. We measured privacy improvements using the Increased Privacy Ratio (IPR), which ranged from 0.76 to 0.89 across batch sizes (e.g., 0.79 for 50 commits, 0.77 for 200, 0.76 for 500, and 0.75 for 1000)and predictive performance using a Random Forest classifier that
remained stable with $F_1$-scores between 0.36 and 0.65 (e.g., $F_1$ = 0.57 for 50 commits, 0.36 for 200, 0.53 for 500, and 0.65 for 1000). These results indicate that LLM based anonymization attains high privacy levels compared to graph anonymization baselines while retaining stable model performance, suggesting that LLMs can serve as a practical foundation for context-aware anonymization.

\subsection{Direct LLM-based Anonymization through zero-shot prompting (RQ1)}
Building on the preliminary study, in this RQ we evaluated whether LLMs can directly anonymize complete commit-level JIT datasets while maintaining both privacy and utility through zero-shot prompting. For this purpose, we have applied \textit{Qwen-3-14B} and \textit{DeepSeek-R1} to the Apache Cassandra dataset~\cite{keshavarz2022apachejit}. We conducted the evaluation of RQ1 using small-scale, open-source models suitable for self-hosting, which ensures reproducibility and provides full control over data handling without relying on proprietary APIs.

We applied a zero-shot prompting approach, where the model was given a prompt describing the dataset schema and the definitions of all metrics (e.g., \textit{la}, \textit{ld}, \textit{nf}, \textit{ndev}, \textit{asexp}), and specified rules for applying Gaussian noise while preserving logical relationships such as \textit{churn = la + ld} and \textit{$aexp \geq arexp \geq asexp$}. The prompt instructed the model to anonymize each batch of 50 commits iteratively, ensuring non-invertible perturbations, preserved class balance, and consistent commit-level structure. This setup guided the LLM to generate privacy-preserving yet utility-consistent data without requiring any in-context examples or fine-tuning.

Each model directly processed the commit data to anonymize and return modified rows, aiming to protect sensitive attributes while maintaining realistic dataset structure. However, both models struggled to perform consistent anonymization across batches. Although the LLMs demonstrated logical reasoning during processing and intermediate explanations, they failed to apply that reasoning effectively when modifying numerical commit metrics. In most cases, the changes were either too small to ensure privacy or inconsistent across commits, highlighting the models’ difficulty in handling numerical perturbations within structured tabular data. These results show that our zero-shot prompting approach alone has not produced reliable anonymization of software metrics, as the lack of statistical control caused inconsistent and insufficient privacy transformations, motivating the structured, cluster-guided LLM-based technique introduced in RQ2.

\subsection{Cluster-guided LLM-based Anonymization (RQ2)}

To address the limitations identified in RQ1, we designed an LLM-based anonymization technique, as shown in Figure~\ref{fig:methodology}, which operates at the cluster level rather than on individual commits. The goal is to make the anonymization process more structured. Instead of treating every commit separately, our approach first groups similar commits into clusters on the basis of QIDs. Each cluster represents a distinct pattern of code change behavior. For every cluster, the LLM produces a set of numerical parameters—such as mixture weights, means, standard deviations, and ratio distributions—that guide how new anonymized values of sensitive attributes should be generated. By working with these aggregated patterns, the model learns what a typical commit looks like in each context and generates realistic, privacy-preserving data that still support accurate JIT defect prediction. 

\subsubsection{Step 1 – Binning Commits}

In the first step, we group similar commits into quantile-based bins so the LLM can reason about statistical patterns within each group instead of treating every commit independently. By this, we ensure that commits with comparable characteristics are grouped together.

We start with ten quasi-identifiers (QIDs): \textit{nf}, \textit{nd}, \textit{ns}, \textit{ent}, \textit{ndev}, \textit{nuc}, \textit{age}, \textit{aexp}, \textit{arexp}, and \textit{asexp}. To form bins, we take each QID column and split its range into twenty parts based on quantiles—that is, we find the values at 5\%, 10\%, 15\%, …, 95\% positions and use them as cut points. This creates roughly equal-sized groups of commits. For example, if \textit{ndev} (number of developers) ranges from 1 to 20, the bins might be (1–2], (2–4], (4–6], etc. Each commit is then linked to one main QID using its unique \texttt{commit\_id}. We calculate a hash of the commit ID and take its remainder (modulus) when divided by the total number of QIDs, ensuring randomness in distribution yet determinism for every commit. In other words, the same commit always maps to the same QID whenever the process is repeated.

\begin{figure*}[t] 
    \centering
    \includegraphics[
        width=\textwidth,
        trim=1cm 11.5cm 0 0, 
        clip
    ]{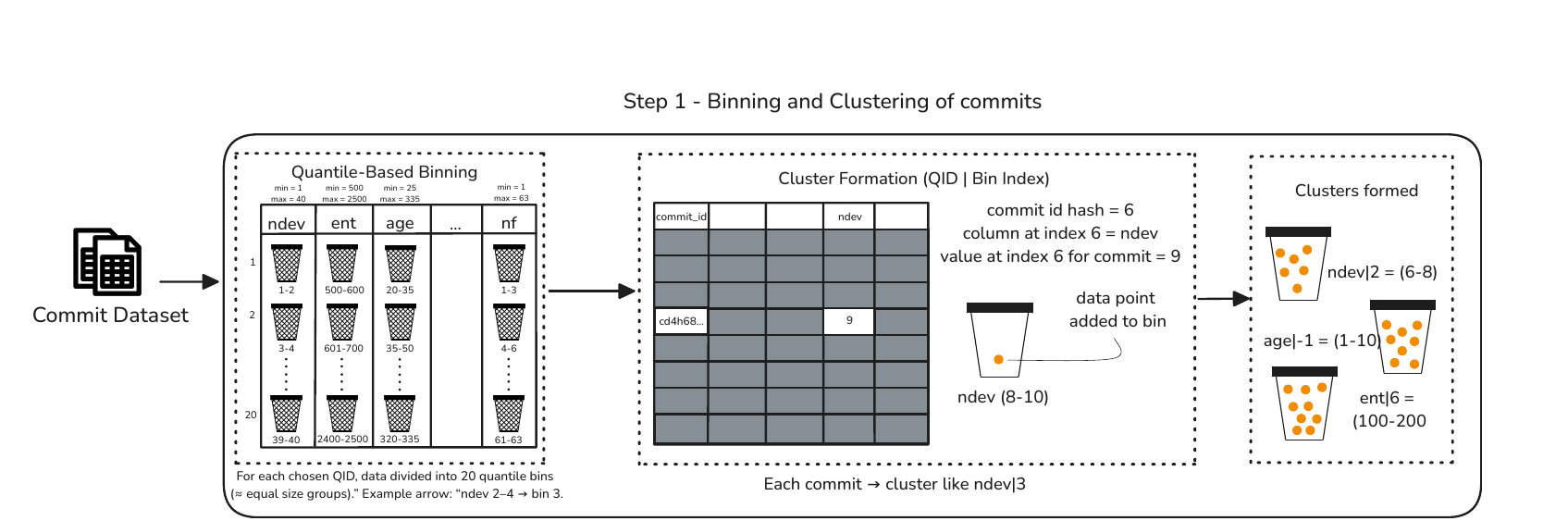}
    
    \caption{Clustering commits via quantile-based QID binning}
    \label{fig:binning}
\end{figure*}

After selecting the primary QID, we look at the commit’s numeric value in that column and see where it falls among the twenty bins. If the value is 7 and the bin edges for that QID are [1, 3, 5, 8, 12], the commit goes into the interval (5–8], which corresponds to bin 3. Commits with missing or out-of-range values are placed into a special bin labeled \texttt{-1}. Finally, each commit receives a cluster label in the format \texttt{<QID>|<bin\_index>}. For example, a commit that hashes to \textit{ndev} and falls into bin 3 is labeled \texttt{ndev|3}, while one that cannot be placed normally is labeled \texttt{ndev|-1}. Figure~\ref{fig:binning} illustrates this process.  

Consider a concrete example: a commit with \texttt{commit\_id = 44a0a3…}, \textit{ndev} = 5, and \textit{age} = 1200 days. The hash of its ID maps it to \textit{ndev}. With bin edges [1, 3, 5, 8, 12], this commit falls into (3–5] and is labeled \texttt{ndev|2}. Another commit may map to \textit{age} and fall into bin 9, labeled \texttt{age|9}. We repeat this process for every commit in the project, ensuring that all commits are deterministically assigned to one of the clusters. These clusters then serve as structured input for Step 2, where we compute descriptive statistics for each cluster to provide context for the LLM during the anonymization process.

\subsubsection{Step 2 – Computing Cluster Statistics}

After forming the bins in step 1, we treat them as \textit{clusters}—each containing commits that share comparable numeric characteristics, such as similar numbers of modified files, developers, or entropy. To describe how commits within each cluster typically behave, we compute concise statistical summaries rather than analyzing individual records. This step gives the LLM structured, group-level context about each cluster’s overall behavior.

For every cluster, we calculate numerical summaries that capture both central trends and variation across its commits. We focus on four key change metrics: lines added (\textit{la}), lines deleted (\textit{ld}), total churn (\textit{la} + \textit{ld}), and the churn ratio (\textit{la}/\textit{churn}). For each of these, we compute descriptive quantiles—the minimum, 5th, 25th, 50th (median), 75th, 95th, and maximum—along with the total number of commits in the cluster. These statistics describe how typical commits in that group behave and how much variation exists between smaller and larger changes, giving the model an overall picture of that cluster’s change patterns. We also record the numeric interval of the cluster’s primary QID, inherited from the quantile boundaries created in Step 1. For instance, a cluster labeled \texttt{ndev|2} may represent commits where the number of developers lies between three and five. Together, these statistics outline what “typical” commits in that cluster look like and how their sizes are distributed.

To illustrate, consider the cluster \texttt{ndev|2} from Step 1, which contains 412 commits with three to five developers. The median churn is 78 lines, with 75 \% of commits below 120 lines and only 5 \% exceeding 210 lines. The median addition ratio (\textit{la}/\textit{churn}) is 0.42, meaning that about 42 \% of modified lines are additions on average. These summaries show that this cluster represents moderately sized, collaborative commits with balanced additions and deletions.

The resulting statistical profiles serve as structured input for the next step, where the LLM uses them to guide anonymization while maintaining the overall data distribution of each cluster.

\subsubsection{Step 3 – LLM-Based Parameter Configuration}

After preparing the statistical summaries, we prompt the LLM to generate anonymization parameters for each cluster. The model’s task is to decide how new values for the sensitive metrics—lines added (\textit{la}) and lines deleted (\textit{ld})—should be produced so that the anonymized data stays realistic while preventing reconstruction of the original values.

We construct a prompt that defines both the expected output format and the logical constraints the model must follow. The prompt consists of two parts:  
\textbf{(1) A system message}, which clearly instructs the model to return only a valid JSON object following a fixed schema; and  
\textbf{(2) A user message}, which provides the cluster-specific data including:
\begin{itemize}
    \item the cluster identifier (\texttt{<QID>|<bin\_index>}),
    \item the range of the primary QID (\texttt{bin\_left}, \texttt{bin\_right}), and
    \item the aggregated statistics from Step~2 (quantiles for \textit{la}, \textit{ld}, \textit{churn}, and the addition ratio \textit{la}/\textit{churn}).
\end{itemize}

The schema inside the system message defines four required fields:
\begin{itemize}
  \item \textbf{cluster\_id} – identifies the cluster being anonymized;
  \item \textbf{churn\_mixture} – specifies how churn (\textit{la} + \textit{ld}) values should vary, using one to three Gaussian components with weights, means, standard deviations, and valid minimum and maximum values;
  \item \textbf{ratio\_beta} – defines a Beta distribution for the ratio\\(\textit{la}/\textit{churn}), ensuring the ratio stays between 0 and 1;
  \item \textbf{constraints} – ensures non-negative integers and the rule $la + ld = churn$.
\end{itemize}

The LLM uses these instructions to infer how new commit values should vary within each cluster. Instead of returning exact replacements, it describes how the data should be regenerated in a realistic way. For instance, it decides how much the total code change (\textit{churn}) can fluctuate around typical values and what proportion of that change is likely to be additions versus deletions. If a cluster mainly contains medium-sized commits with churn between 50 and 200 lines of code, the LLM may suggest that most new commits in that group should fall near 120 lines, with most commits slightly above or below that amount. It also defines how many of those lines are usually added—perhaps around 40–60\%—so the regenerated data preserves the same general balance of additions and deletions. Another cluster with very small commits may instead center around churn values below 50 lines and a higher addition ratio, reflecting mostly additive work.

These numeric patterns give structure to how new values are generated in the next step, ensuring that each cluster retains its realistic behavior. The main benefit of using an LLM here is its ability to automatically learn these patterns from the data summaries rather than relying on hand-tuned parameters. For hundreds of clusters, manually estimating typical sizes, variation ranges, or addition–deletion balances would be both tedious and inconsistent. The LLM, however, captures these relationships contextually, generating coherent numeric settings for each cluster. This makes the anonymization process scalable and adaptive, allowing every group of commits to be anonymized in a statistically consistent yet privacy-preserving way.

\subsubsection{Step 4 – Regenerating Anonymized Commits}

In the final step, we regenerate anonymized commit values for every record using the parameters produced for each cluster. Each commit retains its  non-sensitive information, while the sensitive attributes—lines added (\textit{la}) and lines deleted (\textit{ld})—are replaced with new values that follow the cluster’s statistical behavior. For every cluster, we load its numeric settings describing how total code changes (\textit{churn = la + ld}) and the balance between additions and deletions typically vary. For each commit, its \texttt{commit\_id} serves as a fixed random seed, ensuring deterministic anonymization so the same input always yields identical anonymized values when repeated. Based on these parameters, we generate a realistic total change (\textit{churn}) within the cluster’s range and a proportion (\textit{ratio}) that specifies what fraction of those lines are additions. We then reconstruct the anonymized attributes as \(la = \text{round}(\text{ratio} \times \text{churn})\) and \(ld = \text{churn} - la\). For instance, the same commit from our earlier example in cluster \texttt{ndev|2} (three to five developers) produces a churn of 220 lines and a ratio of 0.35, giving \(la = 77\) and \(ld = 143\), which preserves the pattern of a moderately sized, collaborative change while concealing the original record. This process keeps all commits within a cluster statistically consistent, satisfies logical constraints such as non-negativity and \(la + ld = churn\), and ensures reproducibility across runs while remaining impossible to reverse-engineer. By combining deterministic sampling with LLM-inferred numerical patterns, our technique produces datasets that mirror real development behavior while maintaining both the predictive utility needed for JIT defect prediction and the privacy required for secure data sharing.

\subsection{Cross-LLM Comparison (RQ3)}

RQ3 examines whether our cluster-guided LLM-based technique remains effective when applied to open-source LLMs beyond Qwen-3-14B, which served as the base model in RQ2. The goal is to test the robustness and generalizability of the approach, ensuring that its effectiveness is not restricted to a particular model. To do this, we have repeated the entire anonymization process—including binning, cluster summarization, parameter generation through LLM, and reconstruction—using two additional open-source LLMs with distinct architectures and parameter scales: DeepSeek-R1 and GPT-OSS-20B. 
We selected Qwen-3-14B for its strong performance in quantitative tasks such as metric perturbation \mbox{\cite{yang2025qwen3}}; DeepSeek-R1 (\textasciitilde13B) for its effectiveness in numerical pattern recognition and verifiable reasoning, including math evaluations \mbox{\cite{deepseek-r1-2025}}; and GPT-OSS-20B for its strengths in reasoning tasks and structured output generation \mbox{\cite{agarwal2025gpt}}. These models offer practical computational feasibility for our setting. Larger open models, such as Gemma-3 (27B) and Llama-3 (70B), were excluded due to their significantly higher resource requirements.

Each model received identical statistical summaries for all clusters and independently produced numeric configurations for generating anonymized values. The anonymization and evaluation procedures followed exactly the same setup as in Section~3.4 to ensure fair comparison.  

This design allows us to observe how different LLMs interpret the same statistical input and whether architectural differences influence the privacy–utility balance. Across all datasets, both models generated coherent, stable anonymization outputs that closely matched the privacy and utility levels achieved by Qwen-3-14B. We found that the overall effectiveness of the technique depends on the structured technique itself rather than on the individual LLM, demonstrating that our approach is model-agnostic.

\section{Evaluation}

To assess the effectiveness of our LLM-based anonymization technique, we evaluate both privacy preservation and predictive utility. We conducted all experiments on standard Just-In-Time (JIT) defect-prediction datasets~\cite{keshavarz2022apachejit, mcintosh2018fix}, using the same commit-level metrics defined in prior work~\cite{malik2024towards}. We used a Random Forest classifier~\cite{breiman2001random} as the predictive model, following established practice in defect-prediction research~\cite{mcintosh2018fix, malik2024towards}.

\subsection{JIT Metrics}
We use the commit-level attributes defined in prior work~\cite{malik2024towards, kamei2013large} that capture code churn, developer activity, and change complexity. The metric \texttt{la} represents the number of lines added by the commit, while \texttt{ld} captures the lines deleted. The \texttt{ent} metric quantifies the entropy of changes across modified files, reflecting how evenly the modifications are distributed. Structural impact is captured through \texttt{nf}, the number of files changed; \texttt{nd}, the number of directories affected; and \texttt{ns}, the number of subsystems modified. The metric \texttt{nuc} denotes the number of unique changes within a commit, whereas \texttt{ndev} indicates how many developers previously modified the file. Temporal and experiential aspects are represented by \texttt{age}, which measures the time interval since the file’s last modification; \texttt{aexp}, denoting the author’s overall experience; \texttt{asexp}, the author’s experience within the subsystem; and \texttt{arexp}, the author’s recent experience.


\subsection{Privacy Evaluation}

We adopt the JIT-SDP threat model described in~\mbox{\cite{peters2012privacy, malik2024towards}}.  
The attacker is assumed to know the quasi-identifiers (QIDs), such as \texttt{nf}, \texttt{ndev}, and \texttt{entropy}, but not the sensitive attributes \texttt{la} or \texttt{ld}, and attempts to infer them using QID-based queries.  
Prior studies show that combinations of QIDs can reveal \texttt{la}/\texttt{ld} values~\mbox{\cite{malik2024towards}}.  
The attacker follows the SAD (Sensitive Attribute Disclosure) model~\mbox{\cite{peters2012privacy,malik2024towards}}. To evaluate privacy levels, we simulate this attacker: if the anonymized dataset still allows a QID-only adversary to reconstruct the original \texttt{la}/\texttt{ld} values, anonymization fails; if not, it succeeds.

Privacy is quantified using the \textit{Increased Privacy Ratio (IPR)}~\cite{peters2012privacy}, which measures the proportion of sensitive-attribute queries that yield different results between the original and anonymized datasets. A high IPR indicates that the attacker fails to infer the sensitive attributes (la and ld), while a low IPR indicates that the attacker is close to the true la and ld values.
The IPR is computed as: \textit{IPR = (1$-$(\#Privacy Breaches$/$Total queries)$\times$100}


A ``breach'' occurs when the most common value of a sensitive attribute in the anonymized dataset matches that of the original. We apply the quasi-identifier assumptions (such as the attacker has some knowledge about the value of the quasi-identifier of the entities they wish to identify in the dataset) and equal-frequency binning~\cite{peters2015lace2} to model attacker knowledge.

\subsection{Predictive Utility}
We evaluate predictive utility using a Random Forest classifier~\cite{breiman2001random}, a widely adopted model in software analytics research~\cite{mcintosh2018fix, malik2024towards}. The target variable represents whether a commit is defect-inducing (\texttt{buggy = 1}) or clean (\texttt{buggy = 0}).

All datasets are chronologically divided into an 80\% training and 20\% testing split to mimic real-world project evolution. To enhance robustness, each experiment is repeated with 5 runs, each sampling 500 commits (with replacement). To address class imbalance, we apply the SMOTEENN technique~\cite{batista2004study}, which combines oversampling of minority examples with noise filtering of majority samples.

The Random Forest model is implemented in \texttt{scikit-learn}~\cite{pedregosa2011scikit} with 1,400 estimators, a maximum depth of 100, and \texttt{max\_features = 1}. These hyperparameters follow standard settings in defect-prediction studies to ensure consistent training and evaluation.

\subsection{Evaluation Metrics}
We evaluate model performance using the F1-score, which balances precision and recall by computing their harmonic mean. 
The F1-score is widely adopted in Just-In-Time (JIT) defect-prediction research as a reliable measure of classification performance,~\cite{peters2015lace2,mcintosh2018fix}.


The F1-score thus serves as the primary indicator of model performance throughout our evaluation.

\section{Results}

This section reports the empirical results obtained from our three research questions (RQ1--RQ3). The evaluation focuses on two complementary goals: (i) privacy preservation, quantified using the Increased Privacy Ratio (IPR), and (ii) utility retention, measured through the $F_1$-score of a JIT defect prediction model trained on the anonymized datasets. Across all experiments, higher IPR values indicate stronger protection of sensitive information, while higher $F_1$-scores represent better predictive performance of the model.

\subsection{RQ1 – Direct LLM-based Anonymization through zero-shot prompting}

We first investigate whether LLMs can directly anonymize commit-level data through zero-shot prompting. To test this, we applied QWEN-3-14B and DeepSeek R1 on the Apache Cassandra dataset\\~\cite{keshavarz2022apachejit}, feeding commits in sequential chunks of 50 rows. Each model received raw commit rows containing sensitive metrics, and we explicitly prompted it to anonymize the commits by perturbation. Both models produced coherent outputs that preserved the syntactic integrity of the dataset, but the degree of anonymization remained minimal. In several cases, the models only slightly modified the values, for instance, changing \texttt{la = 12 to la = 11} and \texttt{ld = 4 to ld = 5}, showing that they tried to keep the data realistic rather than significantly perturb it for privacy.

As a result, privacy improvement was marginal. QWEN-3-14B achieved an IPR of 22\% with an F1-score of 70\%, while DeepSeek R1 achieved an IPR of 16\% with an F1-score of 69\%. These IPR values did not even reach the baseline privacy level~1 of 65\%~\cite{peters2015lace2}, which graph anonymization methods typically achieve, indicating that direct LLM-based anonymization failed to provide any practical privacy enhancement. The low IPR values in this experiment confirm that zero-shot prompting alone cannot produce meaningful privacy gains. Both models preferred to keep the data consistent and realistic instead of introducing enough variation to hide sensitive information.

Recognizing these limitations, we developed a cluster-guided LLM-based technique that introduces statistical guidance and adaptive perturbation strength. This structured approach enables the model to generate privacy-preserving yet distribution-consistent anonymizations, overcoming the under-anonymization problem observed in RQ1.

\begin{tcolorbox}[colback=gray!5!white, colframe=gray!60!black, title=\textbf{RQ1 Summary}, left=2pt, right=2pt, boxrule=0.7pt]
\begin{itemize}[leftmargin=1em, noitemsep, topsep=2pt]
    \item Resulting privacy remains low (less than the baseline privacy level~1, i.e., 65\%), with QWEN-3-14B achieving an IPR of 21\% ($F_1 = 70\%$) and DeepSeek R1 achieving an IPR of 15\% ($F_1 = 69\%$).
    \item These findings show that zero-shot prompting under-anonymizes sensitive values, as the models prioritize semantic coherence over disruptive anonymization.
\end{itemize}
\end{tcolorbox}

\subsection{RQ2 – Cluster-Guided LLM-based Anonymization versus Graph Anonymization}

To evaluate the effectiveness of our proposed cluster-guided LLM-based anonymization technique, we compared it with four state-of-the-art graph anonymization methods: Generalization (Gen), k-Degree Anonymity (k-DA), Random Add/Delete, and Random Switch. 
We conducted the evaluation across six widely used JIT defect-prediction projects—Apache Cassandra, Flink, Groovy, Ignite, OpenStack, and Qt~\cite{keshavarz2022apachejit, mcintosh2018fix}. 
All experiments used our LLM-based technique built on the QWEN-3-14B model and followed the same preprocessing and evaluation procedures described in Section~3.4.

To ensure robustness, we repeated each experiment five times for every project. 
We assessed statistical significance using paired Wilcoxon signed-rank tests ($\alpha = 0.05$) for $F_1$-scores, applying the Holm–Bonferroni correction to control for multiple comparisons. 
For privacy scores (IPR), we first applied the Friedman test to verify overall differences among anonymization methods, followed by paired Wilcoxon tests to confirm specific improvements achieved by our LLM-based technique. 
We reported effect sizes using the rank-biserial correlation ($r$) and focused on aggregated results across all baselines to highlight the comparative performance of our approach.

Across all datasets, our LLM-based technique consistently outperformed the graph anonymization baselines in terms of privacy. 
It achieved a mean IPR of 90.2\%, whereas the baselines ranged from 66.0\% (lowest, k-DA) to 79.5\% (highest, Random Add/Delete). 
At the same time, the mean $F_1$-score of our technique (47.5\%) remained comparable to the full range of graph methods, which scored between 41.0\% (lowest, Random Switch) and 47.8\% (highest, Generalization). 
These results confirm that our improvement in privacy did not come at the expense of predictive accuracy.

\begin{figure}[!t]
  \centering
  \includegraphics[
      width=0.7\linewidth,
      trim=0 0.2cm 0 0, 
      clip
  ]{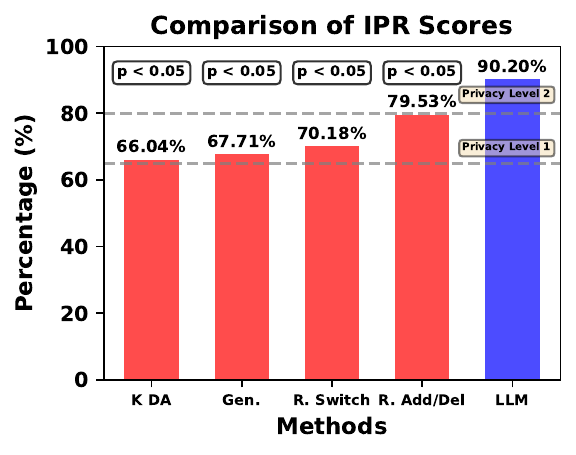}
  
  \caption{Mean IPR across six projects for each anonymization method, showing consistent privacy gains achieved by our LLM-based technique.}
  
  \label{fig:ipr_rq2}
\end{figure}

\begin{figure}[!t]
  \centering
  \includegraphics[
      width=0.75\linewidth,
      trim=0 0.2cm 0 0, 
      clip
  ]{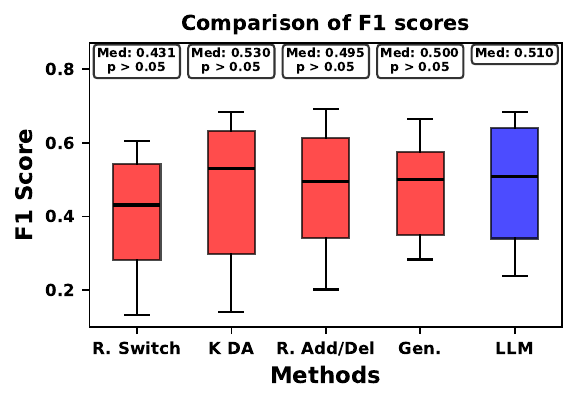}
  
  \caption{Distribution of $F_1$-scores across six projects for each anonymization method. Each box plot shows the median, quartiles, and outliers for all approaches.}
  
  \label{fig:f1_rq2}
\end{figure}

Figure~\ref{fig:ipr_rq2} presents the mean IPR values across all six projects for each anonymization method. 
Our LLM-based technique clearly outperforms all graph baselines, showing IPR gains of +0.106 to +0.241. 
According to the rank-biserial correlation scale~\cite{fritz2012effectsize}, an effect size of $r = 1.0$ represents a \textit{very large} practical impact, indicating consistent and substantial improvement in privacy across all datasets. 
This improvement corresponds to a shift from privacy level~1 (IPR $<80\%$) to level~2 (IPR $\geq80\%$), representing a major enhancement in protection strength.

Figure~\ref{fig:f1_rq2} then summarizes the distribution of $F_1$-scores across the same projects. 
The $F_1$-scores of our LLM-based technique remain within $\pm0.03$ of the baselines and exhibit no statistically significant differences (all adjusted $p>0.05$). 
This stability confirms that the substantial privacy gains achieved by our approach do not compromise predictive performance.

\begin{table}[h]
\centering
\caption{Aggregated mean $F_1$-scores and IPR values across all datasets and anonymization approaches.}

\label{tab:rq2_results}
\begin{tabular}{lcc}
\toprule
\textbf{Approach} & \textbf{Mean $F_1$-Score (\%)} & \textbf{Mean IPR (\%)} \\
\midrule
Generalization & 47.8 & 67.7 \\
k-Degree Anonymity & 47.1 & 66.0 \\
Random Add/Delete & 47.1 & 79.5 \\
Random Switch & 41.0 & 70.2 \\
\textbf{LLM-based} & \textbf{47.5} & \textbf{90.2} \\
\bottomrule
\end{tabular}

\end{table}

Our LLM-based technique achieved mean privacy improvements of $+10.6$ to $+24.1$ percentage points over the graph anonymization baselines, while $F_1$-score differences remained minor ($-0.3$ to $+6.5$ points) and statistically non-significant.

For model utility ($F_1$-scores), paired Wilcoxon tests comparing our LLM-based technique and graph baselines (matched by project) revealed no statistically significant differences. 
Raw $p$-values ranged from 0.031 (Random Switch) to 1.0 (Gen, k-DA), but after Holm–Bonferroni correction all adjusted $p \geq 0.125 > 0.05$. 
Effect sizes ranged from $r = 0.52$ to $r = 1.0$, though the mixed directionality (our method slightly higher for some datasets, slightly lower for others) confirms that predictive performance remains comparable overall.

In our IPR comparisons against the graph-based baselines, the Friedman test showed a statistically significant overall difference among anonymization methods ($p = 0.003$). 
Follow-up paired Wilcoxon tests confirmed that our LLM-based technique achieved significantly higher IPR scores than every baseline, with all $p$-values below 0.05. 
The effect size was maximal ($r = 1.0$), indicating a \textit{very large} practical impact according to the rank-biserial correlation scale~\cite{fritz2012effectsize}. 
These results demonstrate that our proposed technique consistently provides stronger privacy protection than existing graph anonymization methods.

Overall, these findings validate that our LLM-based anonymization technique enhances privacy while maintaining comparable predictive performance, outperforming all graph-based methods in the privacy–utility trade-off. 
The improvement arises from our cluster-guided LLM's context-aware reasoning over feature distributions (e.g., lines added/deleted, churn), enabling adaptive anonymization that produces realistic, privacy-preserving synthetic values.

\vspace{-0.5em}

\begin{tcolorbox}[colback=gray!5!white,colframe=gray!60!black,title=\textbf{RQ2 Summary},left=2pt,right=2pt,boxrule=0.7pt]
\begin{itemize}[leftmargin=1em,noitemsep,topsep=2pt]
    \item Our LLM-based technique achieved a mean IPR of 90.2\%, surpassing all graph baselines (66.0–79.5\%) and exceeding privacy level~2 (IPR~$>80\%$), while maintaining a comparable mean $F_1$ of 47.5\%.
    \item These results confirm that our cluster-guided anonymization maintains predictive accuracy while achieving substantially higher privacy through adaptive, distribution-sensitive perturbation of key metrics (\texttt{la}, \texttt{ld}, churn).
\end{itemize}
\end{tcolorbox}

\subsection{RQ3 – Cross-LLM Comparison}

To assess the robustness and generalizability of our cluster-guided anonymization technique across different LLMs, we replicated the RQ2 experiments using two additional open-source models: GPT-OSS-20B and DeepSeek R1. 
We selected these models to represent variations in architecture, training data, and parameter scale, ensuring that our technique’s performance is not tied to a single LLM (QWEN-3-14B from RQ2). 
All conditions remained identical, including clustering, prompting, preprocessing, and evaluation procedures described in Section~3.4. 
We conducted the comparison across the same six JIT defect-prediction projects—Apache Cassandra, Flink, Groovy, Ignite, OpenStack, and Qt~\cite{keshavarz2022apachejit, mcintosh2018fix}.

Both additional LLMs reproduced the main findings from RQ2: substantial privacy enhancements over all four graph anonymization baselines (Generalization, k-Degree Anonymity, Random Add/Delete, and Random Switch) while maintaining comparable predictive utility. 
This consistency demonstrates that our technique is not dependent on a particular LLM implementation, confirming its broader applicability for privacy-preserving defect prediction. 
Table~\ref{tab:rq3_llm_results} summarizes the aggregated mean $F_1$-scores and IPR values across all projects and baselines.

\begin{table*}[t]
\centering
\caption{Aggregated mean $F_1$-scores and IPR values for cross-LLM comparison across all datasets.}

\label{tab:rq3_llm_results}
\begin{tabular}{lcccc}
\toprule
\textbf{LLM} & \textbf{Mean $F_1$ (Graph-Baseline)} & \textbf{Mean $F_1$ (LLM)} & \textbf{Mean IPR (Graph-Baseline)} & \textbf{Mean IPR (LLM)} \\
\midrule
QWEN-3-14B (RQ2) & 45.7 & 47.5 & 70.9 & 90.2 \\
DeepSeek R1 & 45.8 & 48.5 & 70.9 & 91.1 \\
GPT-OSS-20B & 45.9 & 49.1 & 70.9 & 86.7 \\
\bottomrule
\end{tabular}

\end{table*}

On average, our LLM-based technique improved privacy by +15.8 to +19.3 percentage points relative to the graph baselines, while $F_1$-scores showed minor differences (+1.6 to +3.2 points) without any degradation. 
DeepSeek R1 and QWEN-3-14B produced nearly identical results, while GPT-OSS-20B achieved slightly lower IPR (86.7\%) but the highest $F_1$ (49.1\%), indicating minor model-specific trade-offs.

For predictive utility ($F_1$-scores), paired Wilcoxon tests across projects revealed no statistically significant differences between our technique and graph baselines. 
For privacy preservation (IPR), all LLMs achieved significantly higher scores than the baselines ($p < 0.05$) with a maximal effect size ($r = 1.0$), confirming strong and consistent privacy improvements across all comparisons.

These results affirm the generalizability of our LLM-based technique: privacy improvements remain consistent across diverse LLM architectures, driven by prompts guided by cluster-level statistical summaries that align anonymization with underlying feature distributions.

\begin{tcolorbox}[colback=gray!5!white,colframe=gray!60!black,title=\textbf{RQ3 Summary},left=2pt,right=2pt,boxrule=0.7pt]
\begin{itemize}[leftmargin=1em,noitemsep,topsep=2pt]
    \item All LLMs achieved IPR between 86.7–91.1\%, surpassing graph baselines (70.9\%) with $p < 0.05$ and a very large effect size ($r = 1.0$).
    \item $F_1$-scores remained comparable (47.5–49.1\% vs.~45.7\%), confirming that our LLM-based technique enhances privacy without compromising predictive performance.
\end{itemize}
\end{tcolorbox}

\subsection{Discussion}
The behavior of the LLMs across our experiments reveals how these models interpret and transform structured software-engineering data. In the direct LLM-based anonymization through zero-shot prompting setting (RQ1), the LLMs demonstrated logical reasoning in their responses but failed to meaningfully alter the underlying metrics. This occurs because, during generation, LLMs are guided by an internal sense of plausibility—they prefer values that appear realistic and internally consistent—rather than by explicit statistical objectives. When confronted with numeric commit data, they instinctively maintain coherence across values (for example, keeping the ratio of additions and deletions reasonable) instead of deliberately masking sensitive attributes. Their natural goal is to preserve realism, not to enforce privacy.

The cluster-guided LLM-based technique introduced in RQ2 changes this interaction entirely. Instead of asking the model to directly modify individual commits, it instructs the model to generate parameters that preserve the statistical structure of each commit cluster and later be used to regenerate anonymized values. The model no longer needed to decide what each individual commit should look like; instead, it reasoned about what a typical commit cluster should resemble and produced mixture parameters (weights, means, standard deviations, and ratio distributions) accordingly. This design leverages the LLM’s inherent strength in pattern abstraction and proportional reasoning. The model succeeds here because the anonymization task becomes conceptual rather than computational—requiring reasoning about relationships between variables (e.g., churn and ratio) instead of producing raw numeric noise. Through this reframing, the LLM shifts from data editing to pattern interpretation. It learns that not all commits follow the same behavioral pattern and that different clusters—such as those with high developer activity or large churn—require different anonymization behaviors. The resulting parameter configurations from LLM capture what is typical within each cluster while introducing enough variation in the sensitive attributes to prevent re-identification. This variation helps the anonymized data remain both statistically realistic and useful for prediction. Once these parameters are produced, we use them to regenerate the sensitive values—specifically lines added (\textit{la}) and lines deleted (\textit{ld})—for every commit in each cluster, completing the anonymization process before evaluation.

The same principle explains the method’s strong generalization across LLMs (RQ3). Because anonymization relies on structured statistical descriptions rather than model-specific language behavior, different LLMs trained on distinct corpora produce similar anonymization strategies. The framework provides structure, and the model supplies reasoning—any LLM capable of interpreting quantitative relationships can infer suitable anonymization parameters regardless of architecture or parameter scale.

Moreover, using conventional techniques (graph-based or tabular) is infeasible with our clustering-based approach. Graph techniques rely on removing connections between nodes, but once data is clustered, those connections no longer exist; we no longer have a graph, only groups of similar commits. Tabular techniques shift points toward nearest neighbors, but this would harm bug-prediction utility because it ignores semantic relationships and distorts attribute values.  
Simple random sampling will also perform poorly because defect-prediction data is skewed. Moreover, within small clusters, nearest neighbors are already very similar, so tabular anonymization would barely change the data, leading to weak privacy. In contrast, the LLM can generate new, sufficiently distinct values while preserving semantic relationships.  

A key reason our anonymization avoids exposing sensitive information is that, even though the anonymized dataset preserves cluster-level statistical distributions (e.g., typical churn ranges or typical la/ld ratios), it breaks the instance-level patterns that an attacker needs to reconstruct operational details or developer-productivity indicators. For example, consider a developer who made seven commits last week with the following sensitive pattern: [120, 118, 122, 119, 121, 118, 123] lines added per day. Even if the attacker knows the global distribution of additions for that project (e.g., most commits fall between 50--200 lines), this distribution is not enough to infer that this specific developer consistently adds \texttt{\textasciitilde 120} lines per day. In contrast, our anonymizer regenerates each commit’s la/ld using cluster-level mixture parameters and a commit-specific random seed, producing values such as [65, 102, 154, 93, 141, 87, 110]. These values remain statistically plausible for that cluster (e.g., churn 50--200, ratio 0.35--0.55) but destroy the distinctive daily pattern that reveals the developer’s productivity rhythm.

Overall, the findings show that LLMs serve as effective statistical reasoners when given structured context. Their strength lies in generalizing proportional relationships within data, preserving each cluster’s statistical form while removing individual identity links. This ability to reason conceptually over structured data makes LLMs promising for privacy-preserving analytics.

\section{Implications}


\textbf{Transitioning from Experiments to Deployable Pipelines:} 
Since our cluster-guided LLM-based approach works reliably across different LLMs, it can be integrated as a modular component within existing machine learning pipelines. Teams can reuse our anonymization workflow—clustering, LLM-driven parameter generation, perturbation, and evaluation—without depending on a specific model. This flexibility improves reproducibility in research and scalability in industrial data-sharing scenarios. Our RQ3 results further indicate that different LLMs can replace one another with minimal variation in privacy or utility, enabling organizations to choose models based on computational or regulatory constraints while maintaining consistent anonymization quality.

\textbf{Safer Collaboration and Benchmarking in MSR research:} 
The consistent privacy improvements enabled by the LLM technique facilitate responsible dataset sharing among organizations, researchers, and open-source communities. Since the model maintains utility while protecting privacy, maintainers can release comparable anonymized corpora across multiple projects, supporting cross-organizational benchmarking and replication studies. Adopting standardized privacy levels (e.g., IPR thresholds) can further strengthen reproducibility and fairness in evaluating privacy-preserving techniques.


\section{Threats to Validity}
The threats to the validity of this study include construct, internal, and external aspects. In terms of construct validity, the generation of anonymization parameters by LLMs may depend on prompt formulation, which could influence the resulting data distributions. However, consistent privacy–utility trends across six projects suggest that these effects remain stable. For internal validity, we used a Random Forest classifier following prior JIT defect prediction studies~\cite{mcintosh2018fix, peters2015lace2} and focused on relative changes between anonymized and baseline datasets rather than absolute scores. Multiple randomized runs were conducted to mitigate stochastic variation. Regarding external validity, our evaluation was limited to open-source projects written in Java and Python. While results may vary for other ecosystems or industrial settings, the use of standard JIT metrics and a widely adopted classifier supports the broader applicability of our findings.

\section{Conclusion and Future Work}

In this paper, we presented a cluster-guided LLM-based anonymization technique for Just-In-Time (JIT) defect prediction datasets. By leveraging the contextual reasoning capabilities of Large Language Models, our approach uses the LLM to generate adaptive parameter configurations for anonymization that preserve statistical structure while safeguarding sensitive metrics such as lines added and lines deleted. Empirical evaluation across six software projects showed that our method achieved strong privacy protection (IPR~$\geq$~85\%) with minimal reduction in predictive utility, demonstrating its effectiveness as a context-aware alternative to conventional anonymization techniques. 

For future work, we plan to expand empirical validation to a broader range of open-source and industrial projects to assess scalability and generalizability. We also aim to explore fine-tuning LLMs specifically for software-analytics anonymization to enhance contextual understanding and optimize the privacy–utility balance. Another promising direction is the generation of fully synthetic JIT defect-prediction datasets, enabling privacy-preserving data sharing for the empirical software-engineering community.

\bibliographystyle{ACM-Reference-Format}
\bibliography{references}

\end{document}